# On the Classical Solutions of the Perturbed, Massless Wave Equation with Singular Potential

Ashwin Vaidya

*Department of Mechanical Engineering, University of Pittsburgh, Pittsburgh, PA 15261*

George Sparling

*Department of Mathematics, University of Pittsburgh, Pittsburgh, PA 15260.*

**Abstract**

This paper discusses the solutions to the perturbed wave equation containing a singular potential term in the Lorentzian metric. We present the classical solution to the problem using the separation of variables method for any dimension, $n$. Special solutions are obtained for even $n$'s and properties of these solutions are discussed. Finally, we also consider the solution to the Cauchy problem for the case $n = 2$.

## 1 Introduction

The primary aim of this paper is to discuss the classical[1] structure of the solutions to the perturbed, massless, wave equation

$$\Box \phi + \frac{n(n+2)}{(1+x^2)^2}\phi = 0 \qquad (1)$$

where $\Box$ represents the D'Alembertian operator, $\phi$ is a scalar field and $x = (t, \mathbf{x}) \in \mathbf{M}^n$, $n$ dimensional Minkowski space. The interesting feature of this problem is the external potential term, $\frac{n(n+2)}{(1+x^2)^2}$ with the metric signature (-,+,+,+,...) which gives rise to a singularity as $t^2 \to 1 + x^2 + y^2 + ....$

The existing literature on this subject does not offer concrete examples of equations with singularity and since the solution to the full problem is a rather difficult if not an impossible task, we try and understand the structure of the first order perturbation problem. Yet another motivation for this study was the quantization of the field, $\phi$, which has not been dealt with here and will be the subject of a later paper. The classical results, in particular, the Cauchy Problem, will find use in the quantization of this field. Though our problem is motivated primarily by the physics behind the equations, we choose to study the problem in $n$ dimension since there is considerable interest in the physics community in some dimensions greater than four. The field $\phi$ here can be thought of as a background fluctuation to the wave equation. We wish to understand the behavior of this field particularly in the presence of a singular potential. The motivation for choosing to work with this particular equation is that it provides perhaps the simplest case of a partial differential equation with a singular potential term where calculations can be made in a fairly rigorous manner.

---

[1]The term "classical" here is used in the sense of classical mechanics.

In section 2 we discuss the origins of the perturbed wave equation. We argue here that the powers of the field variables emerge in such a way as to make the field conformally invariant. This invariance property is discussed in section 3. In the following section 4 we discuss the classical solution to equation 1 using the separation of variables method and special properties of solutions are discussed in the case of even dimensions. The final section is devoted to the Cauchy problem for $n = 2$. The discussion is restricted to presenting the final result of the initial value problem; details are omitted due to the tedious nature of the calculations.

## 2 The Perturbed Wave Equation

### 2.1 Origins of the perturbed equation

The perturbed wave equation in $n$ dimensions can be obtained from the Lagrangian density function of the form

$$\mathcal{L} = \int (\frac{1}{2} g^{\mu\nu} \frac{\partial \psi}{\partial x^\mu} \frac{\partial \psi}{\partial x^\nu} - \frac{k(n-2)^2}{2} \psi^{\frac{2n}{n-2}}) \sqrt{\det(g)} d^n x \tag{2}$$

where $k \in \mathbf{R}^+$. Applying the Euler-Lagrange condition to $\mathcal{L}$ yields the $n$-dimensional wave equation, namely

$$\Box \psi + kn(n-2) \psi^{\frac{n+2}{n-2}} = 0. \tag{3}$$

The power of $\psi$ here is chosen so that the field equation remains *conformally invariant* (see section 3). Now, if we choose $\lambda$ to be a scalar and suppose that $\frac{\psi}{\lambda}$ also satisfies the wave equation

$$\Box(\frac{\psi}{\lambda}) + kn(n-2)(\frac{\psi}{\lambda})^{\frac{n+2}{n-2}} = 0. \tag{4}$$

Then, on multiplying by $\lambda$, and choosing $\lambda = k^{\frac{n-2}{4}}$ equation 4 reduces to

$$\Box \psi + n(n-2) \psi^{\frac{n+2}{n-2}} = 0 \tag{5}$$

where we have managed to eliminate the constant $k$. We now obtain a particular solution to equation 5.

**Lemma 1** $\psi_0 = (1 + x^2)^{\frac{2-n}{2}}$ *is a solution of the equation 5.*

**Proof**:
Let us define $\rho := \frac{1}{(1+x^2)}$ then

$$\begin{aligned}
\Box(\rho^r) &= \nabla \cdot (\nabla \rho^r) = \nabla \cdot (-2rx\rho^{r+1}) \\
&= -2nr\rho^{r+1} + 4x^2 r(r+1)\rho^{r+2} \\
&= -2nr\rho^{r+1} + 4r(r+1)\rho^{r+1} - 4r(r+1)\rho^{r+2} = -n(n-2)\rho^{\frac{r+2}{2}}
\end{aligned}$$

if we take, in the final step, $r = \frac{n-2}{2}$ and $\psi_0 = \rho^r$. $\diamond$

We now let $\psi = \psi_0 + \epsilon \phi$ where $\psi_0 = (1 + x^2)^{\frac{2-n}{2}}$ is a solution to the equation 5 proved in Lemma 1. Here $\phi$ can be interpreted as the background fluctuation around $\psi_0$. Then the first order perturbation about this solution yields the equation

$$\Box\phi + \frac{n(n+2)}{(1+x^2)^2}\phi = 0 \tag{6}$$

which is the equation we shall attempt to solve in the following sections. Note that the dimensional dependence of the wave equation appears in the potential term only in the form $n(n+2)$.

## 2.2 Limiting case $n = 2$.

It can be seen from the form of the Lagrangian that the problem is not well defined, as presented above, when $n = 2$. The situation can be remedied by making a simple transformation to the Lagrangian density function, by letting $\psi = \eta^{\frac{n-2}{n+2}}$ in equation 2 and dividing throughout by $(n-2)^2$. The variation of the resulting form of $\mathcal{L}$ for $n = 2$ becomes

$$\Box\chi + 8\frac{k}{a}e^{a\chi} = 0 \tag{7}$$

where $\eta = e^{a\chi}$. It is readily verified that the solution to equation 7 is given by $\chi_0 = \frac{-2}{a}\log(k+x^2)$. Therefore perturbing $\chi$ in the form $\chi = \chi_0 + \epsilon\phi$ yields at $o(\epsilon)$,

$$\Box\phi + \frac{8}{(k+x^2)^2}\phi = 0 \tag{8}$$

which, as can be seen obeys equation 1 for $n = 2$ with $k = 1$.

# 3 Conformal Invariance

Mathematically speaking, we say that a metric $g_{ab}$ is conformally invariant if there is a new metric $\hat{g}_{ab}$ such that $\hat{g}_{ab} = \Omega^2 g_{ab}$ where $\Omega$ is a smooth, positive scalar field. This idea can then be extended to define a conformally invariant field on a manifold (see [5]). A field theory $\psi$ is said to be conformally invariant if $\hat{\psi} = \Omega^k \psi$ also obeys the same field equation. The term $\psi$ is referred to as the *conformal density of weight k*. It should be noted that under rescaling the new metric, $\hat{g}_{ab}$ may not be flat but is said to be conformally flat. But under a suitable choice of weighting factor of $\Omega$, $\hat{g}_{ab}$ will be flat. In general however, since $\hat{g}_{ab}$ is not flat, we need to consider the curvature term, denoted $\mathcal{R}$ to verify invariance. Let us now consider the conformal invariance property of the perturbed wave equation in curved space. It is a well established result that $\Box\psi + \frac{(n-2)\mathcal{R}}{4(n-1)}\psi$ is conformally invariant(see [4]). Therefore it suffices, in our problem, to show that the term $\psi^{\frac{2n}{n-2}}$ is also invariant under conformal transformations. A relatively simple argument for the invariance property of this term is presented. Consider the Lagrangian function (for $n \neq 1$) in curved space given by

$$\mathcal{L} = \int \left(\frac{1}{2}\partial^\mu\psi\partial_\mu\psi - \frac{k(n-2)^2}{2}\psi^{\frac{2n}{n-2}} - \frac{(n-2)\mathcal{R}}{8(n-1)}\psi^2\right)\sqrt{\det(g)}d^n x \tag{9}$$

If we let $g \to \Omega^2 g$ and $\psi \to \Omega^\alpha \psi$ then the relevant term

$$\psi^{\frac{2n}{n-2}}\sqrt{\det(g)} \to \Omega^{\frac{2n\alpha}{n-2}}\psi^{\frac{2n}{n-2}}\Omega^n\sqrt{\det(g)}$$

$$= \Omega^{\frac{2n\alpha}{n-2}+n} \psi^{\frac{2n}{n-2}} \sqrt{\det(g)}$$

$$= \psi^{\frac{2n}{n-2}} \sqrt{\det(g)}$$

for $\alpha = \frac{2-n}{2}$. Therefore for the appropriate choice of weight, $\alpha$, the perturbed wave equation, originating from equation 9, is conformally invariant. Hence for $n = 4$, for instance, $\alpha = -1$.

## 4 Classical Solutions

In this section we shall try to obtain an explicit solution to equation 1 using the separation of variables method. Furthermore, special solutions are obtained in the case of even dimensions. We also derive a recursion relation for these solutions and show completeness.

### 4.1 Solution in $n$ dimensions

Using separation of variables technique we can write

$$\phi = \Sigma_k f_k \left(\frac{-1}{R^2}\right) y_k(x) \tag{10}$$

where $\sqrt{x \cdot x} = R$ is the radial term and $y_k(x)$ (the harmonic term) is a polynomial homogeneous in $x$ of degree $k$ and which obeys the wave equation i.e. $\Box y_k(x) = 0$. Boundary conditions require that $f_k \to 0$ as $R \to \infty$. We also require that each term $f_k \left(\frac{-1}{R^2}\right) y_k(x)$ in the expression above solves the differential equation 1. Substituting for $\phi$ in the perturbed equation we obtain the differential equation

$$4u f_k''(u) - (2n - 8 + 4k) f_k'(u) - \frac{n(n+2)}{(1-u)^2} f_k(u) = 0 \tag{11}$$

where the prime here is used to indicate the derivative with respect to the variable $u$. We then let $f_k(u) = (1-u)^{\frac{n+2}{2}} g_k(u)$. The result of this transformation is the differential equation,

$$g_k'' + \left(\frac{-n+4-2k}{2u(1-u)} - \frac{n+8-2k}{2(1-u)}\right) g_k' - \frac{(n+2)(2-k)}{u(1-u)} g_k = 0 \tag{12}$$

after factoring out $(1-u)^{1+\frac{n}{2}}$.

We recognize this to be *Gauss' differential equation*[2]

$$y'' + \left(\frac{c}{z(1-z)} - \frac{1+a+b}{(1-z)}\right) y' - \frac{ab}{z(1-z)} y = 0 \tag{13}$$

where $a = 2 - k$, $b = \frac{n+2}{2}$, $c = 2 - k - \frac{n}{2}$. The two independent hypergeometric solutions to equation 12 are given by

$$g_1(u) = F[2-k, 1+\frac{n}{2}, 2-k-\frac{n}{2}, u] \tag{14}$$

$$g_2(u) = u^{-1+k+\frac{n}{2}} F[1+\frac{n}{2}, k+n, k+\frac{n}{2}, u] \tag{15}$$

for $|u| < 1$. Gauss' differential equation is known to have twenty-four different solution on the real line. These are also referred to as Kummer's solutions and arise from various transformations of equation 13. The solutions provided above are the first and second Kummer's solutions. Our next attempt is to provide an integral representation for the solutions. On comparison with the twenty-two remaining solutions provided in [2], we realize that

$$g_{12}(u) = (-u)^{\frac{n}{2}}(1-u)^{-1-n}F[-\frac{n}{2}, k-1, k+\frac{n}{2}, \frac{1}{u}] \tag{16}$$

provides the solution to the equation 1. Hence retracing our steps we observe that

$$f_k(u) = (1-u)^{1+\frac{n}{2}} g_{12}(u) \tag{17}$$

where, recall that $u = -\frac{1}{x^2}$ and therefore the solution to the original differential equation is

$$\phi = \int_0^1 p(tx) \frac{(1-t)^{\frac{n}{2}}}{t^2} (t + (\frac{t-1}{1+x^2}))^{\frac{n}{2}} dt. \tag{18}$$

where $p(x) = \sum_k y_k(x)$. That $\phi$ is indeed a solution to the original equation 1 is proved in the proposition below.

**Proposition 1** *The function $\phi$ solves equation 1.*

**Proof**:
We want to show that

$$\Box \phi + \frac{n(n+2)}{(1+x^2)^2} \phi = 0. \tag{19}$$

Let us write

$$f(t,x) = p(tx) \ g(t) \ h(t,x) \tag{20}$$

where

$$g(t) = \frac{(1-t)^{\frac{n}{2}}}{t^2} \ ; \ h(t,x) = (t + (1-t)\rho)^{\frac{n}{2}} \ ; \ \rho = \frac{1}{1+x^2}$$

In this language, the function $\phi$ from above equation 18 is given by

$$\phi := \int_0^1 f(t,x) dt. \tag{21}$$

Then applying the $\Box$ operator to both sides of the above equation gives us

$$\Box \phi = \int_0^1 \Box f(x,t) dt \tag{22}$$

where

$$\Box f(t,x) = \Box(p(tx)g(t)h(t,x)) = p(tx)g(t)\Box h(t,x) + 2g(t)\nabla p(tx) \cdot \nabla h(t,x)$$

We expand this equation below. Note that in our calculations it turns out to be much more convenient to work with the variable $\rho$ rather than $x$. Therefore on simplification we have

$$\Box f(t,x) = ng(t)(1-t)\rho^2 [t + (1-t)\rho]^{\frac{n-4}{2}} [-\rho(-2 + 2\rho + n\rho)p(tx) \tag{23}$$
$$+ \ t(\rho - 1)(-4 + n + 2\rho + n\rho)p(tx) - 2t(t + (1-t)\rho)p'(tx)]$$

In the expression for $\Box f(x,t)$, the term $2g(t)\nabla p(tx)\nabla h(t,x)$ contains a derivative of $p(tx)$ term. So we substitute equation 23 in equation 21 and integrate by parts to obtain an expression purely in terms of $p(tx)$. We write

$$\int_0^1 2g(t)\nabla h(t,x)\nabla p(tx)dt = \int_0^1 \tilde{g}(t)p'(tx)dt = \tilde{g}(t)p(tx)|_0^1 - \int_0^1 p(tx)\tilde{g}'(t)dt$$
$$= -\int_0^1 p(tx)\tilde{g}'(t)dt$$

where

$$\tilde{g}(t) = -2n \quad \rho^2 \quad \frac{(1-t)^{\frac{n}{2}+1}}{t}(t+(1-t)\rho)^{-1+\frac{n}{2}}$$

where we assume that $p(tx)$ vanishes at $t=0$. We can now write

$$\int_0^1 \Box f(t,x) = \int_0^1 p(tx)g(t)\Box h(t,x) + 2\int_0^1 g(t)\nabla p(tx)\nabla h(t,x)$$
$$= \int_0^1 p(tx)g(t)\Box h(t,x) - \int_0^1 p(tx)\tilde{g}'(t)$$

Calculating and substituting for $\tilde{g}'(t)$ and $\Box h(t,x)$ in the above expression and upon simplifying we get

$$\Box f(t,x) = -n(n+2)\rho^2 f(t,x) \tag{24}$$

This proves our claim. ◇

## 4.2 Solution in Even Dimensions

It is observed that the problem simplifies and reveals more interesting features when $n$ is even. The special results we obtain for the case of even dimensions is based on recognition of the fact that the solution to the perturbed equation can be written in the form $\phi = \sum_{r=0}^{\frac{n}{2}} P_r(x)\frac{1}{(1+x^2)^r}$. This is proved in the Lemma below.

**Lemma 2** *The solution to equation 1 can be written in the form*

$$\phi = \sum_{r=0}^{\frac{n}{2}} P_r(x)\frac{1}{(1+x^2)^r} \tag{25}$$

*where $\Box P_r(x) = 0$ for every $r \in [0, \frac{n}{2}]$.*

**Proof**:
The solution to the perturbed equation is given by

$$\phi = \int_0^1 p(tx)\alpha^{\frac{n}{2}}(\frac{1-t}{t^2})^{\frac{n}{2}}dt \tag{26}$$

where $p(tx) = \sum_k p_k(tx)$, $\alpha = (t + \frac{1-t}{1+x^2})$. Hence for $n$ even, $\alpha$ is a polynomial. So using the Binomial Series expansion we have

$$\alpha^{\frac{n}{2}} = t^{\frac{n}{2}}(1 + \frac{1-t}{t(1+x^2)})^{\frac{n}{2}} = t^{\frac{n}{2}} \sum_{r=0}^{\frac{n}{2}} \binom{\frac{n}{2}}{r} \frac{(1-t)^r}{t^r(1+x^2)^r} \quad (27)$$

where $\binom{\frac{n}{2}}{r} = \frac{\frac{n}{2}!}{r!(\frac{n}{2}-r)!}$, the binomial coefficient. Therefore

$$\phi = \sum_{r=0}^{\frac{n}{2}} (\int_0^1 p(tx)(1-t)^{\frac{n}{2}-r} t^{\frac{n}{2}-2-r} dt) \frac{1}{(1+x^2)^r} = \sum_{r=0}^{\frac{n}{2}} P_r(x) \frac{1}{(1+x^2)^r}$$

where $p(tx)$ satisfies the wave equation and therefore, so does $P_r$.  $\diamond$

Therefore, for $n = 2$, $\phi = P_0 + P_1\rho$ and for $n = 4$, $\phi = P_0 + P_1\rho + P_2\rho^2$ etc. This representation is valid for odd $n$ also, although in that case the above representation yields an infinite series. This is not conducive to the kind of calculations that we perform below. In addition, even $n$ suffices for our purposes since our ultimate interest lies in the physically relevant dimension $n = 4$.

We will now show that the equation 1 can be solved directly by taking equation 25 as an ansatz and where the coefficients $P_r$'s are, in fact, not independent but determined by a certain recursion relation and in addition also solve the wave equation. For sake of convenience we shall define the homogeneous operator $H := x \cdot \nabla$ which has the following useful properties :
1. $Hp_n(x) = np_n(x)$
2. The commutation relation, $[\Box, H] = 2\Box$.
3. $\Box H y(x) = (H+2)\Box y(x)$ .

The Theorem below provides proof of the recursion formula satisfied by the coefficients $P_r$. This is followed by a few examples in certain specific dimensions.

**Theorem 1** *The coefficient $P_r$ introduced in the equation 25 obeys a recursion relation given by*

$$P_r = \frac{2(r+1)[2H + (n-2r-4)]P_{r+1}}{(n-2r)(n+2r+2)} \quad (28)$$

**Proof**:
Let us take $\phi$ as in equation 25. Then applying the box operator to both sides of the equation, with the assumption that $\Box P_r = 0$, yields

$$\Box \phi = \sum_{r=0}^{\frac{n}{2}} [P_r(\Box \rho^r) + 2\nabla P_r \cdot \nabla \rho^r]$$

$$= \sum_{r=0}^{\frac{n}{2}} [(4r^2 + 4r - 2nr)P_r - 4rHP_r)\rho^{r+1} - (4r^2 + 4r)P_r\rho^{r+2}].$$

Hence the wave equation becomes

$$0 = \sum_{r=0}^{\frac{n}{2}} [(4r^2 + 4r - 2nr)P_r - 4rHP_r)\rho^{r+1} - (4r^2 + 4r - n^2 - 2n)P_r\rho^{r+2}].$$

$$= \sum_{0}^{\frac{n}{2}-1} 2(r+1)[(2r+4-n-2H)P_{r+1} - (4r^2+4r-n^2-2n)P_r]\rho^{r+2}$$

So it is sufficient that, we have

$$\begin{aligned} 0 &= 2(r+1)[(2r+4-n-2H)P_{r+1} - (4r^2+4r-n^2-2n)P_r] \\ &\Rightarrow \frac{(n-2r)(n+2r+2)}{2(r+1)}P_r = [2H+n-2r-4]P_{r+1} \end{aligned}$$

from which the Theorem follows. $\diamond$

We will now verify by an induction argument that the equation 28 is consistent with the assumption that $P_r$ satisfies the wave equation.

**Proposition 2** *If $P_{\frac{n}{2}}$ obeys $\Box P_{\frac{n}{2}} = 0$ then $\Box P_r = 0$ for any $0 \leq r \leq \frac{n}{2}$.*

**Proof**:

We prove the proposition by a reverse induction argument. It can be seen by concrete calculations that $\Box P_1 = 0, \Box P_2 = 0$, etc. Let us assume that $\Box P_{\frac{n}{2}} = 0$. Then it can easily be shown that all $P_r$'s satisfy $\Box P_r = 0$. Starting from the above theorem, we write

$$P_r = a_1[a_2 H + a_3]P_{r+1}$$

where $a_1 = \frac{2(r+1)}{(n-2r)(n+2r+2)}$, $a_2 = 2$ and $a_3 = n - 2r - 4$. Therefore applying the box operator to both sides of the equation gives us

$$\begin{aligned} \Box P_r &= a_1 \Box [a_2 H + a_3] P_{r+1} \\ &= a_1 a_2 \Box H P_{r+1} + a_1 a_3 \Box P_{r+1} \\ &= [a_1 a_2 (H+2) + a_1 a_3] \Box P_{r+1} \end{aligned}$$

where we have employed the second property of the homogeneous operator in the last equality. Using this last recursion relation in $r$, it is easily seen that $\Box P_r$ must be zero for every $0 \leq r \leq \frac{n}{2}$. $\diamond$

Table 1 provides examples in different dimensions of the recursion relationship obtained in equation 28.

| Dimension, n | Relations | | | |
|---|---|---|---|---|
| 2 | $P_0 = \frac{(H-1)P_1}{2}$ | | | |
| 4 | $P_0 = \frac{HP_1}{6}$ | $P_1 = \frac{(H-1)P_2}{2}$ | | |
| 6 | $P_0 = \frac{(H+1)P_1}{12}$ | $P_1 = \frac{HP_2}{5}$ | $P_2 = \frac{(H-1)P_3}{2}$ | |
| 8 | $P_0 = \frac{(H+2)P_1}{20}$ | $P_1 = \frac{(H+1)P_2}{9}$ | $P_2 = \frac{3HP_3}{14}$ | $P_3 = \frac{(H-1)P_4}{2}$ |

Table I: The recursion relations for $P_r$'s.

## 4.3 Inverse Relations

To establish completeness, we must, in addition to the result of the previous section, also show that the mapping $: P_{\frac{n}{2}} \to \phi$ is surjective. This can be affirmed by checking to see if we can invert the expression in equation 25 above i.e. if we can write each of the $P_r$'s as a function of $\phi$. In this section we shall discuss this inverse relationship for certain specific dimensions. Table 2 summarizes the findings. At the outset, it must be stated that calculations pertaining to inversion becoming extremely tedious with increasing $n$. For this purpose results are shown only up to $n = 6$, although, in principle, we can do so for any even dimension. We shall present our calculations below for dimensions $n = 2$ and $n = 4$. However, unlike in section 4.2, we are unable to obtain a general recursion formula in this case.

1. **Case** $n = 2$

**Lemma 3** *The operator* $(H + 1)^{-1}$ *exists.*

**Proof**:
We define
$$\phi_k = \int_0^1 t^k \phi(tx) dt$$
where $k \geq 0$ ensures convergence of the integral. Then,
$$H\phi_k = \int_0^1 t^k x \cdot \nabla(\phi(tx)) dt = \int_0^1 t^{k+1} \frac{d}{dt}(\phi(tx)) dt.$$

Now integrating by parts we have
$$H\phi_k = \int_0^1 \frac{d}{dt}\left[t^{k+1}\phi(tx)\right] dt - \int_0^1 (k+1)t^k \phi(tx) dt$$
$$= \int_0^1 \frac{d}{dt}\left[t^{k+1}\phi(tx)\right] dt - (k+1)\phi_k$$
$$\Rightarrow (H + k + 1)\phi_k = \phi(x)$$
$$\Rightarrow \phi_k = (H + k + 1)^{-1}\phi(x)$$

The Lemma follows since $\phi$ is well defined and from the definition of $\phi_k$ (by setting $k = 0$ in our case) .
$\diamond$

**Proposition 3** $P_0$ *and* $P_1$ *are invertible and are given by*
$$P_0 = \phi + (H+1)^{-1}(-2\rho\phi)$$
$$P_1 = \rho^{-1}(H+1)^{-1}(2\rho\phi)$$

**Proof**:
We write the equation $\phi = P_0 + \rho P_1$ in the form
$$\rho^{-1}\phi = \rho^{-1}P_0 + P_1 \tag{29}$$

Applying the box operator to both sides of the above equation yields

$$\Box\left(\rho^{-1}\phi\right) = \Box\left(\rho^{-1}P_0 + P_1\right) = \Box\left(\rho^{-1}P_0\right)$$

Carrying the box operator through we have

$$\begin{aligned} LHS &= \Box(\rho^{-1})\phi + \rho^{-1}\Box(\phi) + 2\nabla\rho^{-1}\nabla\phi = \rho^{-1}\Box\phi + 4(H+1)\phi \\ RHS &= \Box(\rho^{-1})P_0 + \rho_0^{-1} + 2\nabla\rho_0^{-1} = 4(H+1)P_0 \end{aligned}$$

Hence equating the two sides and simplifying we get

$$P_0 = \phi + (H+1)^{-1}(-2\rho\phi) \tag{30}$$

and hence it follows that

$$P_1 = \rho^{-1}(\phi - P_0) = \rho^{-1}(H+1)^{-1}(2\rho\phi).$$

$\diamond$

**2. Case $n = 4$**

We follow the same line of arguments as above for this case also

**Lemma 4** *The operator $(H+2)^{-1}(H+3)^{-1}$ exists.*

**Proof**:
Let us define

$$\phi_k = \int_0^1 t^k \phi(tx) dt$$

$$\phi_{k,m} = \int_0^1 \int_0^1 t^k s^m \left(\phi(stx)\right) dtds.$$

Then

$$H\phi_{k,m} = \int_0^1 \int_0^1 t^k s^m H(\phi(stx)) dtds = \int_0^1 \int_0^1 t^{k+1} s^m \frac{d}{dt}(\phi(stx)) dtds$$

using Lemma **??**. Now integrating the right hand side of the last integral by parts we have

$$\begin{aligned} H\phi_{k,m} &= \int_0^1 s^m \phi(sx) ds - (k+1) \int_0^1 \int_0^1 t^k s^m \left(\phi(stx)\right) dtds \\ &= \phi_m - (k+1)\phi_{k,m} \end{aligned}$$

by the definition above. Therefore on simplification

$$\begin{aligned} (H+k+1)\phi_{k,m} &= \phi_m \\ \Rightarrow (H+m+1)(H+k+1)\phi_{k,m} &= (H+m+1)\phi_m = \phi(x) \end{aligned}$$

The last equality follows from Lemma 3. Note that $\phi$ is well defined and using the definitions of $\phi_k$, $\phi_{k,m}$ above it is clearly seen that the operator $(H+2)^{-1}(H+3)^{-1}$ exists (with $k=1$ and $m=2$ in this case). From the pattern of the two Lemma's above, it can be inferred that the Lemma can be generalized for any higher (even) dimension. $\diamond$

**Proposition 4** $P_0$, $P_1$ and $P_2$ are invertible and are given by

$$\begin{aligned}
P_0 &= (H+2)^{-1}(H+3)^{-1}[(H+2)(H+3)\phi - 6\rho(H+1)\phi] \\
P_1 &= (4H+8)^{-1}[\rho^{-1}8(H+3)\phi - 32\phi - 8\rho^{-1}(H+3)P_0 + 8P_0] \\
P_2 &= \rho^{-2}\phi - \rho^{-2}(\phi - 6(H+2)^{-1}(H+3)^{-1}(\rho(H+1)\phi)) \\
&\quad - \rho^{-1}((4H+8)^{-1}[\rho^{-1}8(H+3)\phi - 32\phi - 8\rho^{-1}(H+3)P_0 + 8P_0])
\end{aligned}$$

**Proof**:
We have
$$\rho^{-2}\phi = \rho^{-2}P_0 + \rho^{-2}P_1 + P_2. \tag{31}$$

So
$$\Box(\rho^{-2}\phi) = \Box(\rho^{-2}P_0 + \rho^{-2}P_1 + P_2) = \Box(\rho^{-2}P_0) + \Box(\rho^{-2}P_1)$$

Simplifying the left and right hand sides of the last equation above separately,

$$\begin{aligned}
LHS &= (\Box\rho^{-2})\phi + \rho^{-2}\Box\phi + 2\nabla\rho^{-2}\nabla\phi = \rho^{-1}(8H+24)\phi - 32\phi \\
RHS &= (\Box\rho^{-2})P_0 + (\Box\rho^{-1})P_1 + 2\nabla\rho^{-2}\nabla P_0 + 2\nabla\rho^{-2}\nabla P_1 \\
&= \rho^{-1}(8H+24)P_0 + (4H+8)P_1 - 8P_0.
\end{aligned}$$

Therefore
$$\rho^{-1}(8H+24)\phi - 32\phi = \rho^{-1}(8H+24)P_0 + (4H+8)P_1 - 8P_0 \tag{32}$$

We now apply the box operator a second time on both sides to get

$$\begin{aligned}
\Box^2(\rho^{-2}\phi) &= \Box(\rho^{-1}(8H+24)\phi - 32\phi) \\
&= 24(\Box\rho^{-1})\phi + 24\rho^{-1}\Box\phi + 48\nabla\rho^{-1}\nabla\phi \\
&= 32(H+3)(H+2)\phi - 192\rho(H+1)\phi \\
\Box^2(\rho^{-2}P_0 + \rho^{-2}P_1 + P_2) &= \Box(\rho^{-1}(8H+24)P_0 + (4H+8)P_1 - 8P_0) \\
&= 24(\Box\rho^{-1})P_0 + 48\nabla\rho^{-1}\nabla P_0 + 8(\Box\rho^{-1})HP_0 \\
&\quad + 16\nabla\rho^{-1}\nabla(HP_0) = (32H^2 + 160H + 192)P_0
\end{aligned}$$

So equating the left and right sides above we have
$$P_0 = (H+2)^{-1}(H+3)^{-1}[(H+2)(H+3)\phi - 6\rho(H+1)\phi] \tag{33}$$

Hence using equation 32 we obtain
$$P_1 = (4H+8)^{-1}(\rho^{-1}(8H+24)\phi - 32\phi - \rho^{-1}(8H+24)P_0 + 8P_0) \tag{34}$$

with $P_0$ given as above. Similarly $P_2$ can be obtained in terms of $\phi$, $P_0$ and $P_1$ where $P_0$ and $P_1$ are given above. $\diamond$

| Dimension,n | Relations |
|---|---|
| 2 | $P_0 = \phi + (H+1)^{-1}(-2\rho\phi)$ |
| | $P_1 = \rho^{-1}(H+1)^{-1}(2\rho\phi)$ |
| 4 | $P_0 = (H+2)^{-1}(H+3)^{-1}[(H+2)(H+3)\phi - 6\rho(H+1)\phi]$ |
| | $P_1 = (4H+8)^{-1}[\rho^{-1}8(H+3)\phi - 32\phi - 8\rho^{-1}(H+3)P_0 + 8P_0]$ |
| | $P_2 = \rho^{-2}\phi - \rho^{-2}(\phi - 6(H+2)^{-1}(H+3)^{-1}(\rho(H+1)\phi))$ |
| | - $\rho^{-1}((4H+8)^{-1}[\rho^{-1}8(H+3)\phi - 32\phi - 8\rho^{-1}(H+3)P_0 + 8P_0)]$ |
| $6^a$ | $P_0 = \phi + (H+4)^{-1}(H+3)^{-1}(H+2)^{-1}[12\rho(H+1)(H+2)\phi$ <br> $+ 12\rho^2(H-2)\phi + 24\rho^3\phi]$ |
| | $P_1 = 18(H+2)^{-1}(H+3)^{-1}P_0 + (H+2)^{-1}(H+3)^{-1}$ <br> $[-3\rho^{-1}(H+3)(H+4)P_0 + 3\rho^{-1}(H+3)(H+4)\phi$ <br> $-6(5H+11)\phi + 36\rho\phi]$ |
| | $P_2 = (H+2)^{-1}[2P_1 + 6\rho^{-1}P_0 - 2\rho^{-1}(H+3)P_1 - 3\rho^{-2}(H+4)P_0$ <br> $+ 3\rho^{-2}(H+4)\phi - 18\rho^{-1}\phi]$ |
| | $P_3 = \rho^{-3}\phi - \rho^{-1}P_2 - \rho^{-2}P_1 - \rho^{-3}P_0$ |

Table II: Inverse Relations for $P_r$ for different values of $n$.

[a]Due to the tedious nature of the expressions here, the functions $P_0$, $P_1$, $P_2$ and $P_3$, for the case $n=6$ are expressed as a recursion relation with $P_1$ depending on $P_0$; $P_2$ on $P_1$ and $P_0$; and $P_3$ is expressed in terms of $P_0$, $P_1$ and $P_2$. $P_3$ is expressed in terms of the dependent terms which can be read off from the earlier cell in the table.

## 5 Cauchy Problem

In this section we shall discuss the solution to the Initial Value Problem for the equation 1. The primary motivation for this attempt comes from Segal's outline for quantization (see [1]) which requires us to obtain a *fundamental solution* to the differential equation first. However we restrict ourselves here to the solution to the Cauchy Problem since this is an interesting result in itself and an important aspect of the classical discussion. The discussion, due to the tedious nature of the calculations is restricted to the case $n = 2$. However this gives us a general idea of what can be expected in higher dimensions.

Due to the complexity of the calculations we shall simply provide the final result. It is worth mentioning, however that we solved the problem by analytic continuation into the Euclidean space and then return to the original, Lorentzian space. The solution to the initial value problem for equation 1 in 2 dimensions is given by

$$\begin{aligned} \phi &= \frac{\phi(x-(t-a)) + \phi(x+(t-a))}{2} \\ &\quad - 2\int_{x+(t-a)}^{x-(t-a)} \frac{(t(1+a^2+w^2) + a(x^2 - 2wx - t^2 - 1))}{(1+x^2-t^2)(1-a^2+w^2)^2}\phi(w)dw \\ &\quad - \frac{1}{2}\int_{x+(t-a)}^{x-(t-a)} \frac{(1-x^2+t^2)(1+a^2-w^2) - 4at + 4wx}{(1+x^2-t^2)(1-a^2+w^2)}\varphi(w)dw \end{aligned} \qquad (35)$$

where $\phi(w)$ and $\varphi(w)$ refer to the initial conditions, $\phi|_{t=a} = \phi(w)$ and $\partial_t\phi|_{t=a} = \varphi(w)$. It is easily verified that $\phi$, as given above satisfies the perturbed wave equation.

**Proposition 5** *$\phi$, as given in 35 satisfies equation 1.*

**Proof**

If we define function $\phi(x,t)$ as above then the derivative with respective to $x$ and $t$ respectively are

$$\begin{aligned} \partial_x \phi &= \frac{\phi'(x+t-a)}{2} + \frac{\phi'(x-t+a)}{2} + \frac{2(t-a)\phi(x-t+a)}{(t^2-x^2-1)(1+x^2+t^2-2at+2ax-2xt)} \\ &\quad - \frac{2(t-a)\phi(x+t-a)}{(t^2-x^2-1)(1+x^2+t^2-2at-2ax+2xt)} - \frac{\varphi(x-t+a) - \varphi(x+t-a)}{2} \\ &\quad - a\int_{x+t-a}^{x-t+a} \frac{(w-x)\phi(w)}{(1-t^2+x^2)(1-a^2+w^2)^2}dw + \frac{1}{2}\int_{x+t-a}^{x-t+a} \frac{[4w - 2x(1+a^2+w^2)]\varphi(w)}{(1-a^2+w^2)(t^2-x^2-1)} \\ &\quad + 4x\int_{x+t-a}^{x-t+a} \frac{(t(1+a^2+w^2) + a(x^2-t^2-1) - 2awx)\phi(w)}{(1-t^2+x^2)^2(1-a^2+w^2)^2}dw \\ &\quad + \frac{x}{(1+x^2-t^2)^2}\int_{x+t-a}^{x-t+a} \frac{[4wx - 4at + (1+a^2-w^2)(1-x^2+t^2)]\varphi(w)}{(1+w^2-a^2)}dw \\ \partial_t \phi &= \frac{\phi'(x+t-a)}{2} - \frac{\phi'(x-t+a)}{2} - \frac{2(t-a)\phi(x-t+a)}{(t^2-x^2-1)(1+x^2+t^2-2at+2ax-2xt)} \\ &\quad - \frac{2(t-a)\phi(x+t-a)}{(t^2-x^2-1)(1+x^2+t^2-2at-2ax+2xt)} + \frac{\varphi(x-t+a) + \varphi(x+t-a)}{2} \\ &\quad + \frac{1}{2}\int_{x+t-a}^{x-t+a} \frac{(1+a^2-2at+w^2)\phi(w)}{(1-t^2+x^2)(1-a^2+w^2)^2}dw + \frac{1}{2}\int_{x+t-a}^{x-t+a} \frac{[-4a + 2t(1+a^2-w^2)]\varphi(w)}{(1-a^2+w^2)(t^2-x^2-1)} \end{aligned}$$

$$- \quad 4t \int_{x+t-a}^{x-t+a} \frac{(t(1+a^2+w^2)+a(x^2-t^2-1)-2awx)\phi(w)}{(1-t^2+x^2)^2(1-a^2+w^2)^2} dw$$

$$- \quad \frac{t}{(1+x^2-t^2)^2} \int_{x+t-a}^{x-t+a} \frac{[4wx-4at+(1+a^2-w^2)(1-x^2+t^2)]\varphi(w)}{(1+w^2-a^2)} dw$$

Furthermore, taking two more derivatives, now of the above terms, we get upon calculation and simplification which is not shown here due to the tedious nature of the computations, that

$$- \quad \partial_{tt}\phi + \partial_{xx}\phi = \frac{-4(\phi(x+t-a)+\phi(x-t+a))}{(1+x^2-t^2)^2}$$

$$- \quad 16 \int_{x+t-a}^{x-t+a} \frac{(a^2t+t(1+w^2)-a(1+t^2-x^2+2wx))\phi(w)}{(t^2-x^2-1)^3(1-a^2+w^2)^2}$$

$$+ \quad 4 \int_{x+t-a}^{x-t+a} \frac{(1+a^2-4at+t^2+a^2t^2+w^2(x^2-t^2-1)+4wx-x^2-a^2x^2)\phi(w)}{(t^2-x^2-1)^3(1-a^2+w^2)}$$

$$= \quad -\frac{8\phi}{(1+x^2-t^2)^2}$$

It is also not difficult to see that equation 35 satisfies the initial conditions. The case $\phi|_{t=a} = \phi(w)$ is trivial. For the second condition, upon letting $t = a$, we see that the integral terms vanish. Therefore it is sufficient to consider only the time derivative of the remaining terms which is seen to satisfy the condition $\partial_t \phi|_{t=a} = \varphi(w)$. ◇

The expression for $\phi$ brings out the causal structure of the function, since it is evaluated between the points $x - (t-a)$ and $x + (t-a)$. To know the value of $\phi$ at any point $(x_0, t_0)$, in space-time, we simply need to evaluate the equation 35 between the points $x_0 - (t-a)$ and $x_0 + (t-a)$ inside the past light cone.

# 6  Discussion

We have been successful in unraveling some interesting aspects of the perturbed wave equation. The classical aspects of the field have been explored in some detail, especially in the case of two dimensions. The main contribution of this paper lies in providing a solution to the main problem(equation 1) by a recursive formula, for any even dimension. This allows us to generate the solution to the perturbed problem in terms of the solution to the wave equation. However, explicit results are not possible for odd dimensions. In addition, we also provide a solution to the initial value problem in two dimensions. In fact, the technique used here is valid for any higher dimension (see [3]). The more interesting, and perhaps complicated aspect of the problem, namely, the asymptotics of the field near the singularity is reserved for future study. Along with this the quantization of the field must also be explored. At this stage, however, the paper remains more of mathematical interest. The above mentioned aspects of the problem must be looked at before we can make comments on the physics behind the equations.

# 7  Acknowledgments

This study was part of a masters thesis project at the University of Pittsburgh by Vaidya A. This author would like to thank his advisor and co-author George Sparling for his guidance. He also wishes to acknowledge Bill

Troy for his assistance.